\UseRawInputEncoding
\documentclass[superscriptaddress, notitlepage, reprint]{revtex4-1}
\usepackage[english]{babel}
\usepackage{amsmath,amsthm}
\usepackage{amsfonts}
\usepackage[pdfborder={0 0 0}, colorlinks=true, urlcolor=blue, linkcolor=blue, citecolor=blue]{hyperref}
\usepackage{color}
\usepackage{graphicx}
\usepackage{float}
\usepackage{subfigure}
\usepackage{lipsum}
\usepackage{epstopdf}
\usepackage{dcolumn}
\begin{document}

\title{Quantum Metrology with Coherent Superposition of Two Different Coded Channels}
\author{Dong  Xie}\email{xiedong@mail.ustc.edu.cn}
\affiliation{College of Science, Guilin University of Aerospace Technology, Guilin, Guangxi 541004, People's Republic of China}

\author{Chunling Xu}
\affiliation{College of Science, Guilin University of Aerospace Technology, Guilin, Guangxi 541004, People's Republic of China}

\author{An Min Wang}
\affiliation{Department of Modern Physics, University of Science and Technology of China, Hefei, Anhui 230026, People's Republic of China}

\begin{abstract}
We investigate the advantage of coherent superposition of two different coded channels in quantum metrology. In a continuous variable system, we show that the Heisenberg limit $1/N$ can be beaten by the coherent superposition without the help of indefinite causal order. And in parameter estimation, we demonstrate that the strategy with the coherent superposition can perform better than the strategy with quantum \textsc{switch} which can generate indefinite causal order. We analytically obtain the general form of estimation precision in terms of the quantum Fisher information and further prove that the nonlinear Hamiltonian can improve the estimation precision and make the measurement uncertainty scale as $1/N^m$ for $m\geq2$. Our results can help to construct a high-precision measurement equipment, which can be applied to the detection of coupling strength and the test of time dilation and the modification of the canonical commutation relation.
\end{abstract}
\maketitle

\textit{Introduction}-Quantum metrology mainly involves the use of quantum effects to improve the precision of parameter measurement, which plays a crucial role in the development of basic science and technology~\cite{lab1,lab2,lab3,lab4,lab5,lab6}.
In classical physics, the best metrology precision, known as the quantum shot noise limit (SNL), scales as $1/\sqrt{N}$ with $N$ being the number of resource employed in the measurements. There are two basic ways to beat the SNL and arrive at the Heisenberg limit: one is the parallel entangled-scheme where an entangled state of $N$ probes goes through $N$ maps $U(\varphi)$ in parallel~\cite{lab7}; the other is the sequential scheme where the map $U(\varphi)$ acts on one probe $N$ times.

Quantum \textsc{switch} can offer indefinite causal order, which has potential applications in quantum information, such as  channel
discrimination tasks~\cite{lab8,lab9}. And quantum switches have also been achieved in photonic systems using superpositions of paths for discrete variables~\cite{lab10,lab11,lab12}. Recently, some works showed that quantum \textsc{switch} can increase the quantum Fisher information. Mukhopadhyay \textit{et al.}~\cite{lab13} proposed a novel approach to qubit thermometry using a quantum \textsc{switch} and showed that indefinite causal order can be used as a metrological resource. M. Frey~\cite{lab14} analytically obtained the quantum Fisher information and proved that indefinite causal order can aid quantum depolarizing channel identification.

More importantly, Zhao \textit{et al.}~\cite{lab15} demonstrated that quantum \textsc{switch} can obtain super-Heisenberg scaling based on the sequential scheme in  a continuous variable system. They addressed the comparison with the performances of arbitrary schemes with definite causal order and proved that the optimal scaling is  $1/N^2$.  However, up to now,  the continuous variable quantum \textsc{switch} has not been realized in experiment. An important question arises: can super-Heisenberg scaling be obtained without quantum \textsc{switch}\textbf{?}

In this work, we investigate the advantage of coherent superposition of two different coded channels in quantum metrology. Without quantum \textsc{switch}, the coherent superposition can also obtain super-Heisenberg scaling based on the sequential scheme in a continuous variable system. We also show that the coherent superposition of two different coded channels can perform better than the quantum \textsc{switch}.

Moreover, we analytically obtain the general form of quantum Fisher information and show that nonlinear Hamitonian can improve the estimation precision and make the measurement uncertainty scale as $1/N^m$ for $m\geq2$ based on the sequential scheme.
Our results reveal that the coherent superposition and the nonlinearity can provide an important metrological resource, which can be applied to  conduct high-precision measurement of coupling strength, the gravitational acceleration and the coefficient from the modification of the canonical commutation relations.

\textit{Quantum \textsc{switch}}-Unlike classical physics, the order in which quantum physics allows events to occur is indefinite.  As shown in Fig.~1, based on the sequential scheme, $2N$ black boxes can be accessed, where we consider there are $N$ identical unitary gates $U_1$ and $U_2$.
The quantum \textsc{switch} generates the controlled unitary gate by querying  $U_1$ and $U_2$ gates $N$ times each,
\begin{align}
S(U_1^{\otimes N},U_2^{\otimes N})=|0\rangle\langle0|\otimes U_1^{\otimes N}U_2^{\otimes N}+|1\rangle\langle1|\otimes U_2^{\otimes N}U_1^{\otimes N},
\label{eq:1}
\end{align}
where the first register on the right-hand side of Eq.~(\ref{eq:1}) represents the control qubit.  When
the qubit is in a superposition of $|0\rangle$ and $|1\rangle$,  a coherent superposition
of the two alternative orders $U_1^{\otimes N}U_2^{\otimes N}$ and $U_2^{\otimes N}U_1^{\otimes N}$ (indefinite casual order) can be generated.
\begin{figure}[h]
\includegraphics[scale=0.35]{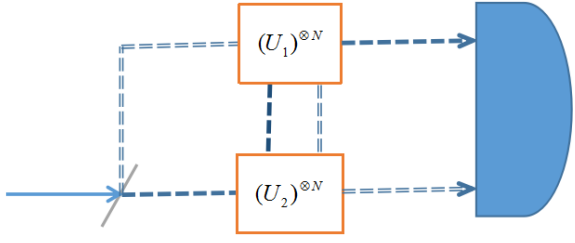}
 \caption{\label{fig.1}Schematic diagram of quantum \textsc{switch}. A control qubit (such as beam-splitter) determines the
order in which quantum operations $U_1^{\otimes N}$ and  $U_2^{\otimes N}$ are applied to the probe state $|\phi\rangle$. Here $U_j^{\otimes N}$ represents that the map $U_j$ acts on the probe $N$ times. When the control is in the superposition state $\frac{|0\rangle+|1\rangle}{\sqrt{2}}$,  there is a superposition of the two orders, generating the output state $ (U_1^{\otimes N}U_2^{\otimes N}|0\rangle|\phi\rangle+U_2^{\otimes N}U_1^{\otimes N}|1\rangle|\phi\rangle)/\sqrt{2}$. }
 \end{figure}
We consider that $U_1$ and $U_2$ are described by
\begin{align}
U_1=e^{-i\theta_1\hat{H}_1}\label{eq:2},\\
U_2=e^{-i\theta_2\hat{H}_2}\label{eq:3}.
\end{align}
In a continuous variable linear system,  we consider that $\hat{H}_1=X$ and $\hat{H}_2=P$. $X$ and $P$  are the conjugate operators, satisfying the canonical commutation relation $[X,P]=i$ ($\hbar=1$ throughout this article).

When the control qubit of the quantum \textsc{switch} is in the state $\frac{|0\rangle+|1\rangle}{\sqrt{2}}$, we can obtain the quantum Fisher information by the formula~\cite{lab16,lab17}
\begin{align}
\mathcal{F_\theta}=4(\langle\partial_\theta\psi|\partial_\theta\psi\rangle-|\langle\partial_\theta\psi|\psi\rangle|^2)
\label{eq:4}.
\end{align}
Here, the final output state is
\begin{align}
|\psi\rangle=(|0\rangle+e^{-i \theta_1\theta_2N^2}|1\rangle)e^{-iN\theta_2P}e^{-iN\theta_1X}|\phi\rangle/\sqrt{2}
\label{eq:4a},
\end{align}
where $|\phi\rangle$ is the initial probe state. Substituting Eq.~(\ref{eq:4a}) into Eq.~(\ref{eq:4}), the corresponding quantum Fisher information are achieved
\begin{align}
\mathcal{F}_{\theta_1}=\theta_2^2N^4+4N^2\langle\phi|\delta^2 X|\phi\rangle\\
\mathcal{F}_{\theta_2}=\theta_1^2N^4+4N^2\langle\phi'|\delta^2 P|\phi'\rangle
\label{eq:4aa},
\end{align}
where $|\phi'\rangle=e^{-iN\theta_1X}|\phi\rangle$ and $\langle\phi|\delta^2 X|\phi\rangle=\langle\phi|X^2|\phi\rangle-|\langle\phi|X|\phi\rangle|^2$. According to the famous Cram$\acute{e}$r-Rao bound~\cite{lab18,lab19}, the estimation precision of $\theta_j$ ($j=1,2$) can be given in the large $\nu$ limit
\begin{align}
\delta\theta_1\approx\frac{1}{\sqrt{\nu}|\theta_2|N^2},\ \delta\theta_2\approx\frac{1}{\sqrt{\nu}|\theta_1|N^2}
\label{eq:5},
\end{align}
where $\nu$ represents the total number of experiments, and we consider that $N\theta_2\gg\langle\phi|\delta^2 X|\phi\rangle$ and $N\theta_1\gg\langle\phi'|\delta^2 P|\phi'\rangle$. We can see that super-Heisenberg scaling $1/N^2$ is achieved.
This recovers the similar results in Ref.~\cite{lab15}.

\textit{Coherent superposition of two different coded channels}-In experiment, there are some difficulties in the realization of continuous variable quantum \textsc{switch}. We propose to use the coherent superposition two different coded channels which does not require quantum \textsc{switch} to obtain high-precision parameter measurement. As shown in Fig.~2, the final output state  $(U_+^{\otimes 2N}|0\rangle|\phi\rangle+U_-^{\otimes 2N}|1\rangle|\phi\rangle)/\sqrt{2}$ is generated by the control qubit, which is in the superposition state $\frac{|0\rangle+|1\rangle}{\sqrt{2}}$. Here, $U_\pm^{2N}$ represents that there are $2N$ identical unitary gates $U_\pm$, which are given by

\begin{figure}[h]
\includegraphics[scale=0.25]{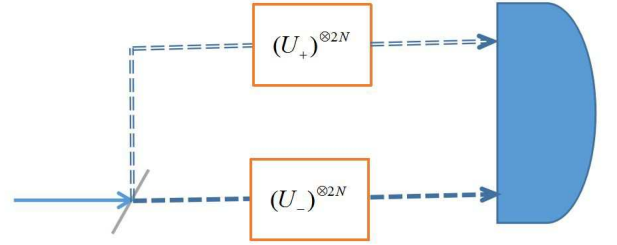}
 \caption{\label{fig.1}Schematic diagram of coherent superposition without quantum \textsc{switch}. When the control is in the superposition state $\frac{|0\rangle+|1\rangle}{\sqrt{2}}$,  there is a superposition of the two different coded channels, generating the output state $(U_+^{\otimes 2N}|0\rangle|\phi\rangle+U_-^{\otimes 2N}|1\rangle|\phi\rangle)/\sqrt{2}$. }
\end{figure}
\begin{align}
U_+=e^{-i\hat{H}_+}=e^{-i(\theta_1\hat{H}_1+\theta_2\hat{H}_2)}\label{eq:6},\\
U_-=e^{-i\hat{H}_-}=e^{-i(\theta_1\hat{H}_1-\theta_2\hat{H}_2)}\label{eq:7}.
\end{align}

In a continuous variable linear system,  we consider that the detail Hamiltonian reads $\hat{H}_\pm=\theta_1X\pm\theta_2P$.
For simplicity's sake, let's just measure $\theta_2$ in the next paragraph.
Using Eq.~(\ref{eq:4}), we can obtain the quantum Fisher information of $\theta_2$,
\begin{align}
\mathcal{F}_{\theta_2}=16N^4\theta_2^2
\label{eq:8}.
\end{align}
The estimation precision of $\theta_2$ can be described as
\begin{align}
\delta\theta_2\simeq\frac{1}{4\sqrt{\nu}|\theta_1|N^2}
\label{eq:9},
\end{align}
where we also consider the large $\nu$ limit.
Comparing Eq.~(\ref{eq:5}) with Eq.~(\ref{eq:9}), we can see that
\begin{align}
\frac{\delta\theta_2|_{\textmd{cs}}}{\delta\theta_2|_{\textmd{qs}}}=\frac{1}{4}
\label{eq:10},
\end{align}
where "cs" denotes the coherent superposition and "qs" denotes the quantum \textsc{switch}. This shows that the coherent superposition without quantum \textsc{switch} can perform better than the case with quantum \textsc{switch} in estimating $\theta_2$ (the results are similar for parameter $\theta_1$).

More importantly, the strategy of coherent superposition of two different coded channels can be realized by a simple Hamiltonian of composite system model
\begin{align}
\hat{H}_{c}=G_1X+G_2\sigma_ZP
\label{eq:11},
\end{align}
where the Pauli operator $\sigma_Z=|0\rangle\langle0|-|1\rangle\langle1|$. Then the output state at time $T$ is given by
\begin{align}
|\psi\rangle_T=e^{-i(G_1X+G_2P)T}|0\rangle|\phi\rangle+e^{-i(G_1X-G_2P)T}|1\rangle|\phi\rangle
\label{eq:12}.
\end{align}
Let us define that $\theta_j=TG_j/2N$ for $j=1,2$. The output state at time $T$ becomes $(U_+^{\otimes 2N}|0\rangle|\phi\rangle+U_-^{\otimes 2N}|1\rangle|\phi\rangle)/\sqrt{2}$. Hence, we show that the composite system can realize the strategy of coherent superposition as shown in Fig.~2. It is much easier than the strategy of quantum \textsc{switch}.

\textit{Nonlinear Hamiltonian}-In a continuous variable nonlinear system,  we first consider a simple nonlinear Hamiltonian, ($H_1=X$, $H_2=P^2$).

In the large $\nu$ and $N$ limit, the estimation precision of $\theta_2$ can be calculated by the above way and the communication relation $[X,[X,P^2]]=-2$,
\begin{align}
\delta\theta_2|_{\textmd{cs}}\approx\frac{3}{16\sqrt{\nu}|\theta_1|^2N^3}, \\
\delta\theta_2|_{\textmd{qs}}\approx\frac{1}{\sqrt{\nu}|\theta_1|^2N^3}
\label{eq:13}.
\end{align}
From the above equations, one can see that both strategies can obtain higher estimation precision with super-Heisenberg scaling $1/N^3$. It demonstrates that nonlinear Hamiltonian can further improve the estimation precision based on the two strategies. And the strategy with the coherent superposition can still perform better than the case with the quantum \textsc{switch}. The radio between $\delta\theta_2|_{\textmd{cs}}$ and $\delta\theta_2|_{\textmd{qs}}$ is
\begin{align}
\frac{\delta\theta_2|_{\textmd{cs}}}{\delta\theta_2|_{\textmd{qs}}}=\frac{3}{16}<\frac{1}{4}
\label{eq:14},
\end{align}
which means that compared with the linear case, the nonlinearity increases the advantage of strategy with the coherent superposition over the strategy with the quantum \textsc{switch} in parameter measurement.

Then, we consider a general nonlinear Hamiltonian, ($H_1=X$, $H_2=P^m$) for integer $m>1$. We can analytically obtain the estimation precision of $\theta_2$ by induction (See Supplemental Material)
\begin{align}
\delta\theta_2|_{\textmd{cs}}\approx\frac{m+1}{2^{m+2}\sqrt{\nu}|\theta_1|^mN^{m+1}}, \\
\delta\theta_2|_{\textmd{qs}}\approx\frac{1}{\sqrt{\nu}|\theta_1|^mN^{m+1}}
\label{eq:15}.
\end{align}
From the above equations, we can see that the super-Heisenberg limit $1/N^{m+1}$ is achieved. It means that  the estimation precision can be further improved as $m$.

After a simple calculation, the radio in the case of the general nonlinear Hamiltonian is described as
\begin{align}
\frac{\delta\theta_2|_{\textmd{cs}}}{\delta\theta_2|_{\textmd{qs}}}=\frac{m+1}{2^{m+2}}
\label{eq:16}.
\end{align}
As $m$ increases, the ratio gets smaller and smaller. It shows that the advantage of strategy with the coherent superposition over the strategy with the quantum \textsc{switch} has been further extended due to the nonlinear Hamiltonian.

\textit{Potential application}-Firstly, we consider that a optomechanical system is formed by
a Fabry-P\'{e}rot cavity with a moving-end mirror. The Hamiltonian of the
system is described as~\cite{lab20,lab21}
\begin{align}
\hat{H}=\omega_c\hat{a}^\dagger \hat{a}+\frac{P^2}{2m}+\frac{1}{2}m\omega_m^2X^2+g\hat{a}^\dagger \hat{a}X
\label{eq:17},
\end{align}
where the last term on the right-hand side of Eq.~(\ref{eq:17}) denotes the radiation pressure on the mirror with coupling strength $g$ and $\hat{a}$ denotes the annihilation operator of the single-mode radiation cavity field with frequency $\omega_c$. $P$ and $X$ are momentum and position operators for the
mechanical oscillator of effective mass $m$, respectively. Here, in order to make better use of the strategy of the coherent superposition,  we consider a low-frequency oscillation with $\omega_m\longrightarrow0$~\cite{lab21a}, leading to that the harmonic potential energy $\frac{1}{2}m\omega_m^2X^2$ is negligible ( like a free particle). And the initial state of the cavity field is given by $(|0\rangle+|1\rangle)/\sqrt{2}$, which generates the coherent superposition. The state at the evolution time $N\tau$ is written as
\begin{align}
|\psi\rangle=(e^{-i\frac{P^2}{2m}N\tau}|0\rangle|\phi\rangle+e^{-i(\omega_c+\frac{P^2}{2m}+gX)N\tau}|1\rangle|\phi\rangle)/\sqrt{2}
\label{eq:18}.
\end{align}
The feasible balanced homodyne detection~\cite{lab22} with the quadrature operator $X=\frac{\hat{a}^\dagger+\hat{a}}{\sqrt{2}}$  is used to measure the coupling strength $g$.
According to the error transfer formula, the uncertainty of $g$ can be derived by
\begin{eqnarray}
\delta^2 g=\frac{\langle X^2\rangle-|\langle X\rangle|^2}{|\partial\langle X\rangle/\partial g|^2}.
\end{eqnarray}
By analytical derivation, we can obtain the estimation precision of $g$ for large $N$
\begin{eqnarray}
\delta^2 g\approx\frac{72m^2(1-\langle U+U^\dagger\rangle^2/8)}{{g^2N^6}|\langle U-U^\dagger\rangle|^2},
\end{eqnarray}
\label{eq:22}
where $\langle U\rangle=\langle\phi|e^{-i(\frac{g^2N^3\tau^3}{6m}+\omega_cN\tau)}e^{-igXN\tau} e^{i\frac{(N\tau)^2P}{2m}}|\phi\rangle$.
Obviously, we can see that the super-Heisenberg scaling $1/N^3$ has been achieved by the  homodyne detection.

Secondly, we consider that a two-level system couples with an oscillator system via dephasing coupling $\sigma_Z\otimes P$, which could be realizable in superconducting qubit-oscillator devices~\cite{lab23,lab24,lab25}. The Hamiltonian of whole system is described as
\begin{align}
\hat{H}\approx G\sigma_ZP+V(X)+\Delta\sigma_Z
\label{eq:23},
\end{align}
where we omit the free Hamiltonian of low-frequency oscillator for the case of strong coupling\cite{lab26} and $V(X)$ denotes the nonlinear potential.
 For example, the nonlinear potential for Duffing system is $V(X)=\beta X^4$~\cite{lab27}. By the similar calculation, we find that the estimation uncertainty of the coupling strength $G$ and the constant $\beta$ are proportional to $1/N^5$.

Thirdly, we consider the gravitational time dilation.  The total Hamiltonian of the system is ~\cite{lab28}
\begin{align}
\hat{H}\approx \frac{P^2}{2m}+g' X (H_0+mc^2)+H_0
\label{eq:23},
\end{align}
where the internal Hamiltonian for this system is $H_0=\sum_{i=1}^M\omega_i\hat{a}_i^\dagger \hat{a}_i$, $m$ denotes the rest mass of the particle in its internal energy
ground state $|0\rangle$, $g'=\frac{g}{c^2}$ and $g$ represents the gravitational acceleration on Earth. Let the initial internal state be the coherent state $(|0\rangle^{\otimes M}+|1\rangle^{\otimes M})/\sqrt{2}$. By the calculation in the above way, the uncertainty of $g$ can also found to be proportional to $1/N^3$. By obtaining the value of $g$, one can check the theory of time dilation in Ref.\cite{lab28,lab29}.

Fourthly, our work can further test the modification of the canonical commutation relations~\cite{lab30}, $[X,P]=i(1+\alpha P^2)$, where the coefficient $\alpha\ll1$.
We consider the strategy of coherent superposition with a general nonlinear Hamiltonian, ($H_1=X$, $H_2=P^m$). The output state is described as
\begin{align}
&|\psi\rangle=\frac{e^{-iN(\hat{H}_1+\hat{H}_2)}|0\rangle|\phi\rangle+e^{-iN(\hat{H}_1-\hat{H}_2)}|1\rangle|\phi\rangle}{\sqrt{2}}\approx\nonumber\\
 &\frac{e^{-i\alpha N^{m+3}}e^{-iN\hat{H}_1}e^{-iN\hat{H}_2}|0\rangle|\phi\rangle+e^{i\alpha N^{m+3}}e^{-iN\hat{H}_1}e^{iN\hat{H}_2}|1\rangle|\phi\rangle}{\sqrt{2}}
\label{eq:12}.
\end{align}
As a result, we find that the scaling of the uncertainty of $\alpha$ is $1/N^{m+3}$. In other words, the nonlinear Hamiltonian can effectively improve the estimation precision of $\alpha$, which will help to test the modification theory of the canonical commutation relations.

\textit{{Conclusion and outlook}}-We have proposed the strategy of coherent superposition of two different coded channels and shown that super-Heisenberg scaling can be achieved in the continuous variable system. In the case of linear Hamiltonian, the strategy of coherent superposition can improve the parameter estimation precision by 4 times compared with the strategy of quantum \textsc{switch}. In the case of nonlinear Hamiltonian,  the enhanced scaling $1/N^{m+1}$ with integer $m>1$ can be obtained. And the nonlinearity further increases the advantage of strategy with the coherent superposition over the strategy with the quantum \textsc{switch} in parameter measurement. Our results provide a high-precision measurement method, which has a potential application in estimating the coupling strength of the optomechanical system and Duffing system. In addition, by enhancing the estimation precision of the gravitational acceleration, the theory of time dilation can be further checked. And we further demonstrate that the nonlinearity can offer a better way to test the modification of the canonical commutation relation.

Adverse conditions, such as uncontrolled environmental disturbances, generally play a detrimental role in quantum metrology.  The further works can be the study of the strategy of coherent superposition of different coded channels in decoherence environment.

\section*{Acknowledgements}

This research was supported by the National Natural Science Foundation of China under Grant No. 62001134 and Guangxi Natural Science Foundation under Grant No. 2020GXNSFAA159047 and National Key R\&D Program of China under Grant No. 2018YFB1601402-2.

\

\section*{Supplementary materials}
\textbf{A. Exponential commutation relation}

Let us first set
\begin{align}
e^{\lambda(A+B)}=e^{\lambda A}e^{\lambda B}e^{\lambda^2 C_2}e^{\lambda^3 C_3}e^{\lambda^4 C_4}... \tag{S1}
\label{eq:S1}.
\end{align}
By differentiating both sides of Eq.~(\ref{eq:S1}) with respect to
$\lambda$ and multiplying it from the right by $e^{-\lambda(A+B)}$, we can obtain
\begin{align}
A+B=A+e^{\lambda A}Be^{-\lambda A}+e^{\lambda A}e^{\lambda B}(2\lambda C_2)e^{-\lambda B}e^{-\lambda A}\nonumber\\
+e^{\lambda A}e^{\lambda B}e^{\lambda^2 C_2}(3\lambda^2 C_3)e^{-\lambda^2 C_2}e^{-\lambda B}e^{-\lambda A}\nonumber\\+e^{\lambda A}e^{\lambda B}e^{\lambda^2 C_2}e^{\lambda^3 C_3}(4\lambda^3 C_4)e^{-\lambda^3 C_3}e^{-\lambda^2 C_2}e^{-\lambda B}e^{-\lambda A}+... \tag{S2}
\label{eq:S2}.
\end{align}
Substituting the known formula $$e^ABe^A=\sum_{i=0}^\infty \frac{1}{i!}[A^{(i)},B]$$ into the above equation, one can obtain
\begin{align}
&0=\sum_{n=1}^\infty \frac{\lambda^n}{n!}[A^{(n)},B]+2\lambda\sum_{m=0}^\infty \sum_{n=0}^\infty \frac{\lambda^{m+n}}{m!n!}[A^{(m)},B^{(n)},C_2]\nonumber\\
&+3\lambda^2\sum_{k=0}^\infty \sum_{m=0}^\infty \sum_{n=0}^\infty \frac{\lambda^{k+m+2n}}{k!m!n!}[A^{(k)},B^{(m)},C_2^{(n)}, C_3]+4\lambda^3\nonumber\\
&\sum_{k'=0}^\infty\sum_{k=0}^\infty \sum_{m=0}^\infty \sum_{n=0}^\infty \frac{\lambda^{k'+k+2m+3n}}{k'!k!m!n!}[A^{(k')},B^{(k)},C_2^{(m)},C_3^{(n)}, C_4]\nonumber\\
&+... \tag{S3}
\label{eq:S3}.
\end{align}
in which,
\begin{align}
&[A^{(0)},B]=B, [A^{(n+1)},B]=[A,[A^{(n)},B]]\nonumber\\
&[A^{(k)},B^{(m)},C_2^{(n)}, C_3]=[A,[A^{k-1},B^{(m)},C_2^{(n)}, C_3]].\nonumber
\end{align}

Based on the above equation, we can obtain the detail form of $C_n$, such as,

\begin{align}
C_2=-\frac{1}{2}[A,B], C_3=\frac{1}{3}[B,[A,B]]+\frac{1}{6}[A,[A,B]]\tag{S4}
\label{eq:S4}.
\end{align}
In principle, the formula can be derived by Eq.~(\ref{eq:S3}). However, one can not obtain the general simplified form of $C_n$  for any $n$. It can't be written in a compact form.

Fortunately, when $A=X$ and $B=P^m$, we can obtain a simple results by following induction
\begin{align}
&C_2=-\frac{1}{2}[A,B],\nonumber\\
&C_3=\frac{1}{6}[A^{(2)},B]],\nonumber\\
&C_4=-\frac{1}{24}[A^{(3)},B]],\nonumber\\
&C_5=\frac{1}{120}[A^{(4)},B]]\tag{S5}
\label{eq:S5}...
\end{align}
According to the above equations, we can get
\begin{align}
C_n=(-1)^{(n-1)}\frac{1}{n!}[A^{n-1},B]\nonumber\\
=(-i)^{(n-1)}\frac{m!}{n!(m-n+1)!}P^{m-n+1}\tag{S6}
\label{eq:S6}.
\end{align}
For $\lambda=1$, the exponential commutation relation is described as
\begin{align}
e^{(X+P^m)}=e^{ X}e^{ P^m}e^{\sum_{n=2}^{m+1} C_n}\tag{S7}
\label{eq:S7}.
\end{align}

When $A=P^m$ and $B=X$, we can get
\begin{align}
C'_2=\frac{1}{2}[B,A],\nonumber\\
C'_3=-\frac{1}{3}[B^{(2)},A]],\nonumber\\
C'_4=\frac{1}{8}[B^{(3)},A]],\nonumber\\
C'_5=-\frac{1}{30}[B^{(4)},A]]\tag{S8}
\label{eq:S8}.
\end{align}
 By induction, in this case, $C'_n$ is described as
\begin{align}
C'_n=(-1)^{n}\frac{n-1}{n!}[B^{n-1},A]\nonumber\\
=-(-i)^{(n-1)}\frac{m!(n-1)}{n!(m-n+1)!}P^{m-n+1}\tag{S9}
\label{eq:S9}.
\end{align}
For $\lambda=1$, the exponential commutation relation is described as
\begin{align}
e^{(X+P^m)}=e^{ P^m}e^{ X}e^{\sum_{n=2}^{m+1} C'_n}\tag{S10}
\label{eq:S10}.
\end{align}
\textbf{B. Measurement with the strategy of the coherent superposition}

Utilizing Eq.~(\ref{eq:S6}), we can express the output state in the case of the coherent superposition
\begin{align}
e^{-i2(\theta_1X+\theta_2P^m)N}|0\rangle|\phi\rangle+e^{-i2(\theta_1X-\theta_2P^m)N}|1\rangle|\phi\rangle\nonumber\\
=e^{-i2N\theta_1X}e^{-i2N\theta_2P^m}e^{\sum_{n=2}^{m+1}(-2Ni)^n\theta_1^{n-1}\theta_2C_n}|0\rangle|\phi\rangle+\nonumber\\
e^{-i2N\theta_1X}e^{i2N\theta_2P^m}e^{\sum_{n=2}^{m+1}-(-2Ni)^n\theta_1^{n-1}\theta_2C_n}|1\rangle|\phi\rangle\tag{S11}
\label{eq:S11}.
\end{align}
Then, the quantum Fisher information can be achieved by $\mathcal{F_\theta}=4(\langle\partial_\theta\psi|\partial_\theta\psi\rangle-|\langle\partial_\theta\psi|\psi\rangle|^2)$. As a result, we obtain the general formula

\begin{align}
\mathcal{F}_{\theta_2}|_{\textmd{cs}}=&4|\langle-2iNP^m+m\theta_1(-2Ni)^2P^{m-1}/2+\nonumber\\
&...+(-1)^m\theta_1^m(-2Ni)^{m+1}\frac{-1}{m+1}\rangle|^2,\tag{S12}
\label{eq:S12}
\end{align}
where $\langle\bullet\rangle=\langle\phi|\bullet|\phi\rangle$. For $N\gg\theta_1\langle P\rangle$,
\begin{align}
\mathcal{F}_{\theta_2}|_{\textmd{cs}}\approx\frac{2^{2(m+2)}\theta_1^{2m}N^{2(m+1)}}{(m+1)^2}.\tag{S13}
\label{eq:S13}
\end{align}

\textbf{C. Measurement with the strategy of the quantum \textsc{switch}}

Utilizing Eq.~(\ref{eq:S10}), we can express the output state in the case of the quantum \textsc{switch}
\begin{align}
e^{-iN\theta_1X}e^{-iN\theta_2P^m}|0\rangle|\phi\rangle+e^{-iN\theta_2P^m}e^{-iN\theta_1X}|1\rangle|\phi\rangle=\nonumber\\
e^{-iN\theta_1X}e^{-iN\theta_2P^m}(|0\rangle|\phi\rangle+\nonumber\\
e^{-imP^{m-1}\theta_1\theta_2N^2+...
+(-i)^mN^{m+1}\theta_1^m\theta_2}|1\rangle|\phi\rangle)\tag{S14}
\label{eq:S14}.
\end{align}

Then, the quantum Fisher information can be achieved by the similar way
\begin{align}
&\mathcal{F}_{\theta_2}|_{\textmd{qs}}=2(|\langle NP^m\rangle|^2\nonumber\\
&+|\langle-iNP^m-im\theta_1N^2P^{m-1}+...+(-i)^m\theta_1^mN^{m+1}\rangle|^2\nonumber\\
&-|\langle-i2NP^m-im\theta_1N^2P^{m-1}+...+(-i)^m\theta_1^mN^{m+1}\rangle|^2).\tag{S15}
\label{eq:S15}
\end{align}

 For $N\gg\theta_1\langle P\rangle$, the above equation can be simplified as
\begin{align}
\mathcal{F}_{\theta_2}|_{\textmd{cs}}\approx\theta_1^{2m}N^{2(m+1)}.\tag{S16}
\label{eq:S16}
\end{align}
\end{document}